\documentclass[twocolumn,pra,aps,superscriptaddress]{revtex4-1}
\usepackage[latin9]{inputenc}
\usepackage{mathtools}
\usepackage{amsbsy}
\usepackage{amssymb}
\usepackage{amstext}
\usepackage{bm}
\usepackage[unicode=true,pdfusetitle,
bookmarks=false,
breaklinks=false,pdfborder={0 0 1},backref=false,colorlinks=true,allcolors=blue]
{hyperref}
\hypersetup{
	bookmarksnumbered=false,bookmarksopen=false}
\makeatletter

\@ifundefined{textcolor}{}
{%
	\definecolor{BLACK}{gray}{0}
	\definecolor{WHITE}{gray}{1}
	\definecolor{RED}{rgb}{1,0,0}
	\definecolor{GREEN}{rgb}{0,1,0}
	\definecolor{BLUE}{rgb}{0,0,1}
	\definecolor{CYAN}{cmyk}{1,0,0,0}
	\definecolor{MAGENTA}{cmyk}{0,1,0,0}
	\definecolor{YELLOW}{cmyk}{0,0,1,0}
}

\newcommand{\beq}{\begin{equation}}
\newcommand{\eeq}{\end{equation}}
\newcommand{\beqa}{\begin{eqnarray}}
\newcommand{\eeqa}{\end{eqnarray}}

\usepackage{amsmath}
\usepackage{graphicx}
\usepackage{amssymb}
\usepackage{txfonts,color}
\makeatother
\begin{document}
	
\title{Active Learning for the Optimal Design of Multinomial Classification in Physics}

\author{Yongcheng Ding}
\email{jonzen.ding@gmail.com}
\affiliation{Department of Physical Chemistry, University of the Basque Country UPV/EHU, Apartado 644, 48080 Bilbao, Spain}
\affiliation{ProQuam Co., Ltd., 200444 Shanghai, China}

\author{Jos\'e D. Mart\'in-Guerrero}
\email{jose.d.martin@uv.es}
\affiliation{IDAL, Electronic Engineering Department,  ETSE-UV, University of Valencia, Avgda. Universitat s/n, 46100 Burjassot, Valencia, Spain}

\author{Yujing Song}
\affiliation{International Center of Quantum Artificial Intelligence for Science and Technology (QuArtist) \\ and Department of Physics, Shanghai University, 200444 Shanghai, China}

\author{Rafael Magdalena-Benedito}
\affiliation{IDAL, Electronic Engineering Department, ETSE-UV, University of Valencia, Avgda. Universitat s/n, 46100 Burjassot, Valencia, Spain}

\author{Xi Chen}
\email{chenxi1979cn@gmail.com}
\affiliation{Department of Physical Chemistry, University of the Basque Country UPV/EHU, Apartado 644, 48080 Bilbao, Spain}

\date{\today}

\begin{abstract}	
Optimal design for model training is a critical topic in machine learning. Active Learning aims at obtaining improved models by querying samples with maximum uncertainty according to the estimation model for artificially labeling; this has the additional advantage of achieving successful performances with a reduced number of labeled samples.  We analyze its capability as an assistant for the design of experiments, extracting maximum information for learning with the minimal cost in fidelity loss, or reducing total operation costs of labeling in the laboratory. We present two typical applications as quantum information retrieval in qutrits and phase boundary prediction in many-body physics. For an equivalent multinomial classification problem, we achieve the correct rate of 99\% with less than 2\% samples labeled. We reckon that active-learning-inspired physics experiments will remarkably save budget without loss of accuracy.

\end{abstract}

\maketitle
\section{Introduction}
Machine learning (ML) has conquered intricate tasks in the past decade~\cite{go,sc}. A critical obstacle to applying ML is that collecting sufficient labeled data is both time-demanding and resource-consuming. Consequently,  model training requires some sort of optimization, aiming at deriving a well-trained model, even making use of numerous unlabeled data, as it is common real-world problems. For now, physicists also complete quantum tasks, study properties of quantum systems, and design physics experiments with ML algorithms~\cite{asnqe,qmbann,bhml,anotheral,njpoverlap,qmcnn,dqmbd,vnnoqs,cnss,dpcnn,ciqereview,fedorov}. Its most utilized branch, so-called reinforcement learning (RL)~\cite{rltextbook}, has naturally shown its capability in quantum control~\cite{bukovprx,qslrd,xinwangnpj,zhouepl,bergli,prati,nevennpj,rbrl,sadowski,shersonnpj,ueda,paparelle,yaodapn,kweksp,dingstadl,ustcdrl}. It is also related to quantum information retrieval by controlling the measurement process~\cite{proj,bsanders}, which has been extended to a quantum version~\cite{qrl} for optimal measurement control~\cite{pancho,panchito}. To minimize the cost of measurements, one has to design the optimal strategy for quantum information retrieval, which fits the framework of active learning (AL). 

The key hypothesis of AL is that a model trained on a subset of adequately selected samples to be labeled can achieve a similar performance as the one trained with all samples labeled~\cite{nlp,settles}. It is verified that AL achieves an accurate binary classification of quantum information by selecting the quantum states with the maximal information for labeling by measurements~\cite{yongcheng2020}. In this paradigm, the cost of labeling is the fidelity loss induced by measurement, which depends on the measurement strength and feedback. The application of AL is not bounded to quantum information retrieval, where we have shown the trade-off between extracted quantum information (model refining) and fidelity loss (cost of labeling). Recently, it is also employed to assist experimental control~\cite{zhaicpl,hush,appaml}, computational physics~\cite{jchemphys,alloy,eprm,zhaiprr}, quantum machine learning~\cite{mamd,qnal,bpe}, etc., attaining convincing performance as well. Based on these facts, we conclude that most of the physics problems can be efficiently studied by AL, if they can be equivalently represented by classification problems. Accordingly, the cost of labeling is no longer limited to the fidelity loss in quantum information retrieval, but extended to the operation cost that reduces the uncertainty of samples by experimental protocols, including doing numerical simulations or physics experiments for analyzing the most informative patterns queried by AL.

In this work, we present typical applications of AL algorithms on classification problems in physics. We focus on multinomial cases, where different sampling strategies are no longer equivalent. Starting from a nontrivial extension of the framework proposed in~\cite{yongcheng2020}, we propose quantum information retrieval in qutrit systems, allowing an experimental implementation with six entangled photons. Another example comes from many-body physics, where AL manages to predict the boundaries of the exotic phases with less than 2\% of samples labeled by phase detections. We also verify that AL guarantees the optimal model training by querying the most informative samples,  so that the rate estimation can hardly be improved without new artificially labeled samples.  In summary,  we reckon that the introduction of AL algorithms would remarkably reduce the cost of physics experiments.

\section{Active Learning Theory}
\begin{figure}
\includegraphics[width=7.8cm]{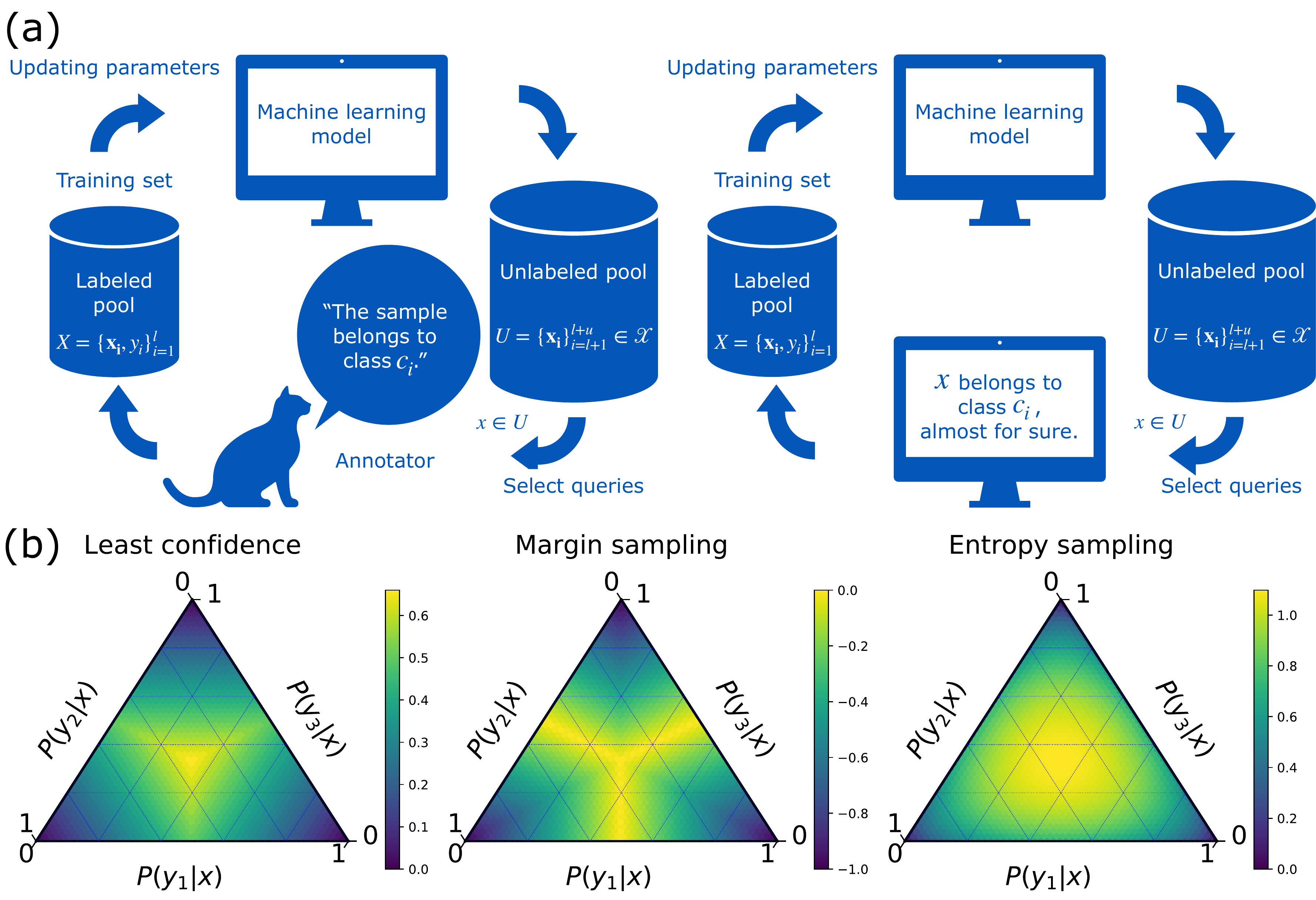}
\caption{\label{fig:scheme}(a) The schematic diagrams of pool-based AL cycle (on the left) and SSL cycle (on the right). (b) The querying behavior of least confidence, margin sampling, and entropy sampling in a triple classification problem is illustrated by heatmaps. The most informative query regions radiate from the centers. The label with the highest probability locates near the corner, where the opposite edge shows the lowest probability for the rest of classes.}
\end{figure}

Given a set of labeled samples $X=\{\mathbf{x_i},y_i\}_{i=1}^l$, where the inputs $\mathbf{x_i}\in \mathcal{X}$, being $\mathcal{X}$ defined in parameter space of $d$-dimension $\mathbb{C}^d$. We consider a classification problem, looking for a model $\theta$ that predicts the output, corresponding to the class $y_i\in\mathcal{C}=\{c_1,\cdots,c_m\}$ for an $m$-class problem. Another set of unlabeled samples $U=\{\mathbf{x_i}\}_{i=l+1}^{l+u}\in\mathcal{X}$ is required for the application of AL,  having $u\gg l$, i.e., one has much more samples as candidates to be labeled than those in the labeled pool. AL works differently from supervised learning, which trains the model with a fixed labeled set.  AL follows an iterative procedure based on adding the most informative sample from $U$ to the training set $X$ in each iteration, in order to improve the model performance; the goal is to find a satisfying model quickly with only a few samples in $U$ being labeled manually. Now the critical question is which samples contain the most information that should be selected and labeled? Normally the selection is performed by models, suggesting the samples with maximal uncertainty about the outcome. Thus, labeling these samples provides more information, benefiting the learning process by adding them to the training set $X$. Although AL updates its training set iteratively like semi-supervised learning (SSL), they are not the same concept that is often confused. Here we clarify the difference by a schematic diagram [c.f. Fig.~\ref{fig:scheme}(a)]. SSL is also based on a labeled training sample $X$ and a pool of unlabeled samples $U$. The main difference between AL and SSL is that the latter instead of selecting the most informative sample of maximal uncertainty,  makes use of a model $\theta$ that selects the sample $\mathbf{x_i}\in U$ with maximal certainty of its class $y_i$, labeling it as $y_i$ since one can assume that the probability of wrong labeling is minimum. Even though AL can be combined with SSL, e.g., training the model by SSL after an initial selection of labels by AL, it does not significantly benefit the learning procedure.

Various approaches allow us to evaluate the uncertainty for sorting the samples in $U$ and making the decision about which candidate should be transferred to $X$ after labeling. There are two widely practiced strategies for this goal, which are the so-called uncertainty sampling (USAMP) and query-by-committee (QBC)~\cite{costsen}. USAMP only uses a single model for selecting samples according to the estimator~\cite{usamp}. QBC employs a committee that consists of several models to select the samples with the minimal consensus, measured by voting entropy or Kullback-Leibler divergence~\cite{qbc}. For realizing USAMP in multiple class problems, we introduce three strategies as follows.  The Least Confidence criterion 
is based on labeling the sample with the least confidence according to the prediction
\begin{eqnarray}
x_{\text{LC}}&=&\underset{x}{\text{argmax}}\left[1-P_{\theta}\left(\hat{y} \arrowvert x\right)\right],\nonumber\\
\hat{y}&=&\underset{y}{\text{argmax}}\left[P_{\theta}\left(y \arrowvert x\right)\right],
\end{eqnarray}
where $\hat{y}$ is the most probable class according to the probabilistic classification model $\theta$. One notices that the model only considers the most probable label $\hat{y}$, losing information of the other labels. A natural extension is that the first and second most probable class labels $\hat{y_1}$ and $\hat{y_2}$ contain more information, which might be better if a corresponding criterion is constructed. Therefore a more informative approach called margin sampling is given as
\begin{equation}
x_{\text{M}}=\underset{x}{\text{argmin}}\left[P_\theta\left(\hat{y_1}|x\right)-P_\theta\left(\hat{y_2}|x\right)\right],
\end{equation}
inspired by the truth that one can easily classify the samples separated by large margins, and accordingly the most ambiguous sample is given by small margins~\cite{margin}. Thus, the sample with the smallest margin is the most informative one since knowing its true label from the human annotator contributes the most value to discriminate among $m$ classes. Obviously, this criterion can be further extended by taking information from all classes, with information entropy being quantified by Shannon's theory~\cite{entropy} as
\begin{equation}
x_{\text{E}}=\underset{x}{\text{argmax}}\left[-\sum_iP_\theta\left(\hat{y_i}|x\right)\log P_\theta\left(\hat{y_i}|x\right)\right].
\end{equation}
The information entropy measures the amount of information required for representing a piece of given information, which is usually considered as an assessment of the uncertainty, being widely used in all aspects of ML.  The implicit relationship among these measures leads to divergent query behaviors [c.f. Fig.~\ref{fig:scheme}(b)] when dealing  with problems of multiple classes. This might cause varying performance after the training, especially when the class distribution of samples is biased. 

In the case of QBC,  voting entropy can be defined by
\begin{equation}
x_{\text{VE}}=\underset{x}{\text{argmax}}\left[-\sum_i \frac{V(y_i)}{\tilde{V}}\log\frac{V(y_i)}{\tilde{V}}\right],
\end{equation}
where $V(y_i)$ denotes the votes from the committee of the size $\tilde{V}$ for the label $y_i$. Although QBC shows its advantage on performance over USAMP in the previous binary quantum information classification task~\cite{yongcheng2020}, we focus on USAMP in this work since different query strategies can be investigated by multinomial classification problems in physics with only a single model, saving massive computational resources.

\section{Informational Retrieval in Qutrits}
According to quantum information theory, the carriers with a higher dimension of Hilbert space encode more information than a qubit as a two-level system. Here we consider a nontrivial extension to the binary classification problem proposed in~\cite{yongcheng2020}, substituting the qubits by qutrits. Information is extracted from the qutrits by quantum measurement for classification, where the cost for labeling is defined as the fidelity loss for consistency.
\subsection{Measurement for labeling}
A qutrit is a three-level system that allows independent transition between levels, which can be constructed by logical basis states selected from a biphoton system. We only consider an arbitrary qutrit wave function $|\Psi\rangle=c_1|0\rangle+c_2|1\rangle+c_3|2\rangle$ for labeling by quantum measurement, without focusing on its physical realization. The qutrit is in the superposition of three orthogonal bases, where each of the bases denotes a class. Once a qutrit is queried, we aim at retrieving its population of each basis by quantum measurement, labeling it as the class of maximal population as a human annotator. As the weak measurement is introduced as an extension of von Neumann measurement for extracting information from a quantum system without collapsing it, we employed the weak measurement in our framework in order to reduce fidelity loss. The weak measurement for labeling binary class can be simplified by coupling an ancilla qubit as a pointer to the qubit. The expectation $\langle\hat{\sigma_z}\rangle$ can be estimated by weak values from the pointer, classifying the sample qubit $\langle\hat{\sigma_z}\rangle>0$ for class 0 and $\langle\hat{\sigma_z}\rangle<0$ for class 1. However, the expectation on the $Z$ direction of the spin-1 operator $\langle\hat{S_z}\rangle=|c_1|^2-|c_3|^2$ does not carry enough information for picking out the basis of the maximum population anymore. For example, the qutrit can be either class 2 if $P(c_1)=0.1$, $P(c_2)=0.7$, and $P(c_3)=0.2$, or class 3 if $P(c_1)=0.3$, $P(c_2)=0.3$, and $P(c_1)=0.4$, with the same expectation $\langle\hat{S_z}\rangle=-0.1$. The trick for equivalently evaluating the maximum population is no longer available in qutrit, hence the need to retrieve the amplitude of each basis or the diagonal elements of the system density matrix. Instead of quantum state tomography, we employ the recent (exact) direct reconstruction scheme~\cite{rqdm} to retrieve the information in qutrits by weak measurement, allowing arbitrary coupling strength without approximation.

Weak measurement protocol usually consists of two steps: coupling the quantum system to a quantum pointer for a new system, then performing a projective measurement on the pointer. Instead of a Gaussian pointer of position, we use ancilla qubits as two-level pointers. We couple the ancilla qubit to the qutrit following an interaction Hamiltonian
\begin{equation}
H_I(t)=g(t)\hat{A}\otimes Y,
\end{equation}
where $g(t)$ is the time-dependent coupling, $\hat{A}$ is the observable to be weakly measured, and $Y$ is the spin-$1/2$ Pauli operator on the $Y$ direction. The qutrit is coupled to the pointer after $t_0$, described by the unitary operator
\begin{equation}
U=\mathcal{T}\exp\left[-i\int_0^{t_0}g(t)\hat{A}\otimes Ydt\right]=\exp\left(-i\theta\hat{A}\otimes Y\right),
\end{equation}
where $\theta$ should be sufficiently small for approximating $U=\mathbb{I}-i\theta\hat{A}\otimes Y$. Additionally, the scheme guarantees that
\begin{equation}
U=\left(\mathbb{I}-\Pi\right)\otimes\mathbb{I}+\Pi\otimes\exp\left(-i\theta Y\right)
\end{equation}
is valid for all coupling $\theta$ when the observable is a projection operator. Exact direct reconstruction of the density matrix can be realized by an arbitrary strength of weak measurement, where $\theta=\pi/2$ provides a maximally strong measurement. Two ancilla qubits A and B are needed as pointers for retrieving quantum information carried in the qutrit. The initial system is described by
\begin{equation}
\rho_{\text{ini}} = \rho\otimes|0\rangle_A\langle0|\otimes|0\rangle_B\langle0|,
\end{equation} 
where the element of the density matrix $\rho_{jk}$ can be expressed in the orthonormal basis of the $d$-dimensional Hilbert space ($d=3$ for qutrit, with orthonormal basis be $|0\rangle$, $|1\rangle$, and $|2\rangle$). We couple the projection operator $\Pi_{a_j}=|a_j\rangle\langle a_j|$ to the $Y$ direction of the first ancilla qubit, giving the evolution operator
\begin{equation}
U_{A,j}=(\mathbb{I}-\Pi_{a_j})\otimes\mathbb{I}_A\otimes\mathbb{I}_B+\Pi_{a_j}\otimes\exp(-i\theta_AY_A)\otimes\mathbb{I}_B,
\end{equation}
then followed by coupling $\Pi_{b_0}=|b_0\rangle\langle b_0|$ to the second ancilla qubit
\begin{equation}
U_B=(\mathbb{I}-\Pi_{b_0})\otimes\mathbb{I}_A\otimes\mathbb{I}_B+\Pi_{b_0}\otimes\mathbb{I}_A\otimes\exp(-i\theta_BY_B),
\end{equation}
where $|b_0\rangle=d^{-1/2}\sum_j|a_j\rangle$ is independent of $j$ or $k$. The system is evolved by
\begin{equation}
\rho_{\text{couple},j}=U_BU_{A,j}\rho_{\text{ini}}U_{A,j}^\dag U_B^\dag,
\end{equation}
being ready for information extraction by projectively measurement (post-selection) of the qutrit in basis $|a_j\rangle$ and corresponding measurement on ancilla qubits. For labeling the qutrit sample, we need the population of each basis, i.e., the diagonal elements of the qutrit density matrix. The following expectation value
\begin{equation}
\langle\hat{A_A}\hat{A_B}\rangle_{j,k}=\text{Tr}\left[\left(\Pi_{a_k}\otimes\hat{A_A}\otimes\hat{A_B}\right)\rho_{\text{couple},j}\right]
\end{equation}
can be measured for the task. Specifically, the diagonal elements can be exactly estimated:
\begin{equation}
\rho_{jj}=16N_{AB}^2\langle\Pi_{1A}\Pi_{1B}\rangle_{j,k},
\end{equation}
by defining $\Pi_{1A,B}=|1\rangle_{A,B}\langle1|$ the projection operators on the the Pauli $Z$ operator's eigenbasis $|1\rangle$ of ancilla qubits, being independent of $k$. The factor $N_{AB}=d/(4\sin\theta_A\theta_B)$ can be inferred by normalizing the density matrix after all elements are estimated, if the coupling strength $\theta_{A,B}$ are unknown. Different from the weak measurement scheme in~\cite{yongcheng2020}, the estimation of $\rho_{jj}$ is precise for any $\theta$ since the ancilla qubits here are not Gaussian pointers with low accuracy of position. Meanwhile, signals should be amplified by a proportion of $\sin^4\theta$ for reconstructing the density matrix. Although the statistical errors are inevitable for an arbitrary measurement strength, the errors become more relevant because of the biquadratic weak signal. Similarly, the statistical error with certain $\theta_A$ and $\theta_B$ can be reduced by $1/\sqrt{n}$ if $n$ copies are given to measure the correlator $\langle\Pi_{1A}\Pi_{1B}\rangle$ for retrieving $\rho_{jj}$. For evaluating the labeling cost, we calculate the fidelity between the initial qutrit density matrix $\rho$ and the density matrix after coupling $\tilde{\rho}$
\begin{equation}
F=\left[\text{Tr}\left(\sqrt{\sqrt{\tilde{\rho}}\rho\sqrt{\tilde{\rho}}}\right)\right]^2,
\end{equation}
where $\tilde{\rho}$ is defined as tracing over the ancilla qubits
\begin{equation}
\tilde{\rho}=\text{Tr}_{A,B}\left(\rho_{\text{couple},j}\right)=\sum_i\left(\mathbb{I}_{\text{qutrit}}\otimes\langle b_i|\right)\rho_{\text{couple},j}\left(\mathbb{I}_{\text{qutrit}}\otimes |b_i\rangle\right),
\end{equation}
with $|b_i\rangle$ be the orthonormal basis of sub-Hilbert space of the ancilla qubits.

\subsection{Problem formulation}
Now we demonstrate AL with a triple classification problem, in which the quantum information to be retrieved is encoded in qutrits. We realize the qutrits by biphotons, which the logical basis states are selected by the polarization of photons as follows
\begin{equation}
|0\rangle=|HH\rangle,~|1\rangle=\frac{1}{\sqrt{2}}\left(|HV\rangle+|VH\rangle\right),~|2\rangle=|VV\rangle,
\end{equation}
where $H$ and $V$ denote the horizontal and vertical polarization mode of a photon, respectively. Arbitrary states can be prepared with an adequate experimental setup, allowing efficient manipulation of qutrits. Considering the advantages of qutrits in quantum communication, we use the remote state preparation (RSP) protocol~\cite{rsp}, whose experimental setup is also compatible with the initial weak measurement framework. Alice wants to send (remotely prepare) a qutrit state $|\Psi\rangle=c_1|0\rangle+c_2|1\rangle+c_3|2\rangle$ to Bob. To this aim, Alice starts with preparing a maximally entangled qutrit state
\begin{equation}
|\Psi_\text{ent}\rangle=\frac{1}{\sqrt{3}}\left(|0\rangle_a|0\rangle_b+|1\rangle_a|1\rangle_b+|2\rangle_a|2\rangle_b\right),
\end{equation}
then followed by measuring her qutrit $a$ with the projection $|\Psi^*\rangle=c_1^*|0\rangle+c_2^*|1\rangle+c_3^*|2\rangle$. Bob will accept his qutrit $b$ if he gets a positive signal from Alice, claiming that she managed to project her qutrit $a$ to $|\Psi^*\rangle$. Otherwise, the qutrits are discarded, being replaced by a new iteration of RSP. Alice understands that an arbitrary projection should satisfy $|c_1|^2+|c_2|^2+|c_3|^2=1$, deciding to test Bob with the following task; it is assumed that $c_i$ are functions of $\mathbf{x}\in\mathcal{X}$, where Bob has no information of $c_j(\mathbf{x})$. Alice prepares $l+u$ qutrits according to the functions $c_j(\mathbf{x_i})$, where $\mathbf{x_i}$ are sites in the parameter space of $d$-dimension. Alice asks Bob to classify these qutrits with minimal fidelity loss, inferring the relation between the sites $\mathbf{x_i}$ and the class $c_j$. Alice allows Bob asking for sending him an arbitrary qutrit among $l+u$ samples, allowing all possible quantum operations on his qutrit as well.

\begin{figure}
\includegraphics[width=7.8cm]{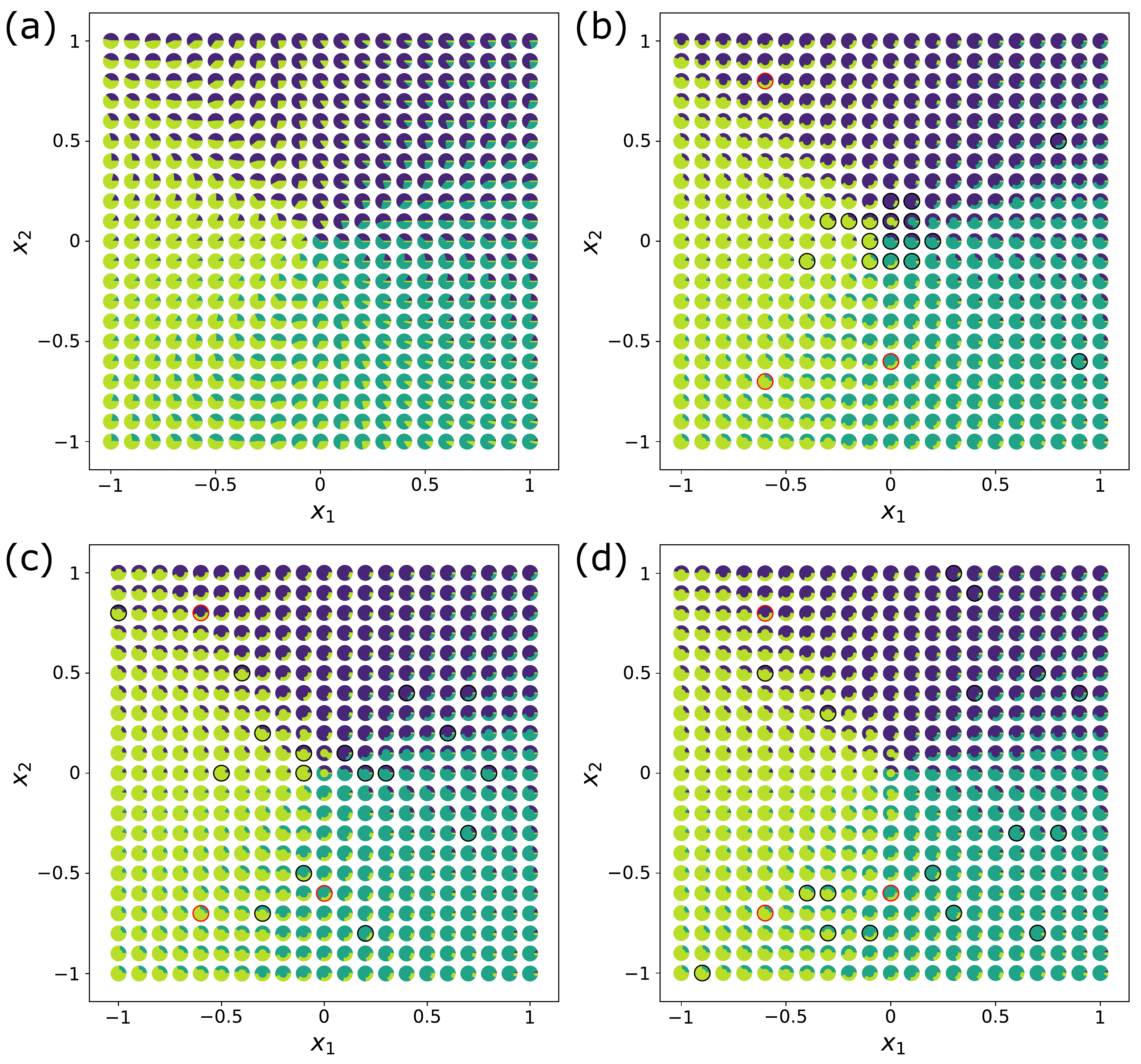}
\caption{\label{fig:alicebob}(a) The lattice prepared by Alice consists of $21\times21=441$ qutrits for triple classification. The qutrit states are demonstrated by pie charts with proportions be $|c_1|^2$, $|c_2|^2$, and $|c_3|^2$. (b), (c), and (d) USAMP protocols with the query strategies be least confidence, margin sampling, and entropy sampling, respectively. Bob initializes a logistic regression model by three oracles provided by Alice (circled by red). Qutrits queried by the model are circled by black and covered by smaller circles in class colors. We find out that queries behaviors are different even if we employ a model initialized by the same training set. Meanwhile, all sampling strategies lead to a quick model convergence and satisfying estimation.}
\end{figure}

\subsection{Numerical simulation}
In Fig.~\ref{fig:alicebob} (a), Alice prepares a quantum state in a lattice of $21\times21=441$ qutrits, encoding the map $x_1,x_2\rightarrow c_1,~c_2$, and $c_3$, which are the amplitudes of logical basis states. Bob receives the correct labels of three qutrits from Alice, belonging to different classes for initializing the classification model. With the linearly-separable assumption, Bob selects logistic regression because it is probabilistic and simple. Bob labels a qutrit by direct reconstruction of its density matrix, randomly retrieving a diagonal element, labeling it as class $j$ if $\rho_{jj}>0.5$. Otherwise, Bob retrieves another diagonal element, which carries enough extra information for labeling the qutrit. In this way, Bob selects the candidate among unlabeled samples based on uncertainty, defined by various criteria as least confidence, margin sampling, and entropy sampling, adding them to the training set for tuning the model [see Fig.~\ref{fig:alicebob} (b-d)]. The model converges to a nearly 90\% correct rate with less than 5\% labeled samples. We notice that different sampling strategies give similar estimations but different distribution patterns of sampled qutrits. For example, least confidence tends to select qutrits located in the middle of the parameter space, while qutrits near the borders of two classes are more likely to be queried by margin and entropy sampling. Here we neglect the probability of samples being wrongly labeled due to the statistical error of $\langle\Pi1_A\Pi_B\rangle_{j,k}$, which can be reduced by asking for more copies of the qutrit candidate.

\begin{figure}
\includegraphics[width=7.8cm]{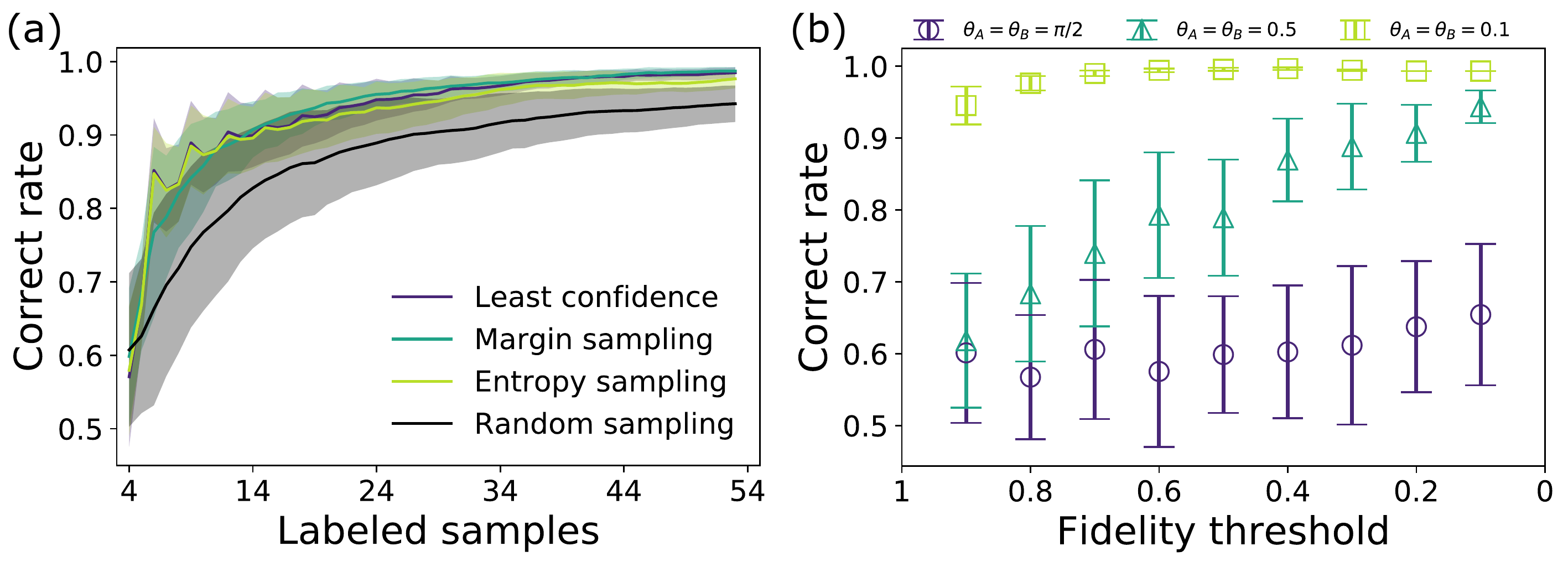}
\caption{\label{fig:fidelity}(a) Mean correct rates of triple classification model with random sampling (baseline), least confidence, margin sampling, and entropy sampling as different sampling strategies. Confidence intervals are filled by transparent colors, denoting a standard deviation based on $200$ numerical experiments. (b) Mean correct rates of triple classification model with margin sampling. Each qutrit is labeled by measurements of different strength $\theta_A$ and $\theta_B$. Parameters remain the same as in the previous subfigure.}
\end{figure}

In order to study different strategies of AL quantitatively, we define the cost of labeling in the direct reconstruction of density matrices with arbitrary coupling strength by average fidelity loss. We bound the number of labeled samples in the training set or fidelity loss to compare the three strategies. In the latter case, we stop labeling qutrits once the fidelity of the system reaches the threshold. Different from~\cite{yongcheng2020}, we no longer take $n$ copies as variables into consideration, i.e., all samples are labeled correctly. Thus, it is trivial that a weaker coupling strength $\theta_i$ should lead to better performance without the trade-off of requiring more copies, otherwise increasing the probability of contaminating the training set with incorrect labels. In Fig.~\ref{fig:fidelity}(a), the result shows that all three strategies outperform random sampling (RS) as the baseline, indicating that AL works with the application of the logistic regression model on this dataset, which we should not take it for granted (see Appendix.~\ref{sec:AL}). Although we can hardly separate these strategies, we may choose margin sampling for its smoother curve and stabler performance in this case, according to the numerical experiment presented in Fig.~\ref{fig:fidelity}.

\section{Phase Transition In Many Body Physics}
The cost of labeling refers to the fidelity loss induced by extracting quantum information from the samples in the previous section, which can be extended to other definitions in various scenarios. For example, the exotic features of many-body physics can be studied by numerical simulations or laboratory experiments, which are both time-demanding and resource-consuming. We employ AL to study a phase transition estimation problem, aiming to efficiently classify the multiple phases in the model.

\subsection{Magnets with geometrical frustration}
\begin{figure}
\includegraphics[width=7.8cm]{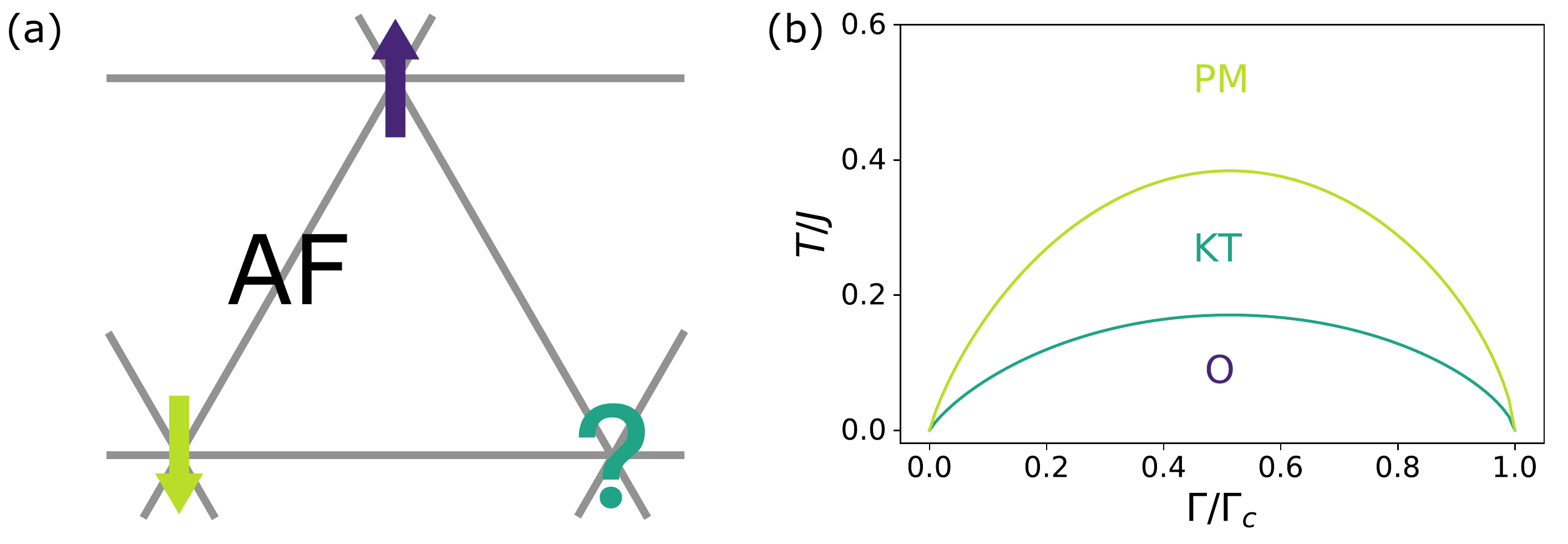}
\caption{\label{fig:tiaf}(a) An intuitive illustration of geometrical frustration that stems from the antiferromagnetic Ising model on a triangular lattice, where spins are aligned opposite to neighbors for minimizing the energy. A spin is frustrated once the other two align antiparallel since either orientation gives the same energy. (b) Phase diagram of the model under transverse field with boundaries,  eq.~\eqref{eq:boundary}, given by previous researches~\cite{ying2005,ying2006} . There exists a clock (ordered) phase since the frustration effect occurs to each spin, giving a ground state of sixfold degeneracy. A critical (Kosterlitz-Thouless,  KT) phase floats between the paramagnetic phase and clock phase, where KT transitions occur at the phase boundaries.}
\end{figure}

The antiferromagnetic Ising model on a triangular lattice (TIAF) under transverse field has the quantum Hamiltonian
\begin{equation}
H = J\sum_{\langle i,j\rangle}\sigma_i^z\sigma_j^z-\Gamma\sum_i\sigma_i^x,
\end{equation}
which might be the simplest model for realizing geometrical frustration [c.f.~Fig.~\ref{fig:tiaf}(a)]. There exists an extended critical phase in transverse field TIAF, being separated by two Kosterlitz-Thouless (KT) transitions from sublattice ordered phase on one side and paramagenetic (disordered) phase. One can study its quantum dynamics with analytic methods~\cite{ying2005,ying2006}, allowing quantitative prediction of the phase diagram at both arbitrary finite temperature and transverse field strength, giving the phase boundary of the KT phase and paramagnetic phase
\begin{equation}
\label{eq:boundary}
\frac{T_2}{J}=b\frac{\Gamma}{\Gamma_c}\ln^\nu\left(\frac{\Gamma_c}{\Gamma}\right),
\end{equation}
where $b=0.98$ is the numerical constant fixed by renormalization, $\nu\approx2/3$ is the 3D XY exponent, and $\Gamma_c=1.65J$ is the critical strength of transverse field. Correspondingly, we have the boundary of the KT phase and ordered phase by the substitution $T_1=(4/9)T_2$~[c.f.~Fig.~\ref{fig:tiaf}(b)]. The numerical values of $b$ and $\Gamma_c$ are suggested by previous density matrix renormalization group (DMRG) and Monte Carlo studies of the system~\cite{isakov}.

Meanwhile, we are aware of the fact the statistical properties of the system can be derived from its partition function $Z=\text{Tr}\exp(-\beta H)$, where $\beta=1/(k_{\text{B}}T)$ ($k_{\text{B}}=1$ for simplicity). According to the Suzuki-Trotter theorem~\cite{suzuki}, we can express the partition function of the quantum Hamiltonian in terms of a stacked classical Ising model with the reduced Hamiltonian
\begin{equation}
H_{2+1} = \sum_{\langle i,j\rangle,k}K_{ij}s_{ik}s_{jk}-\sum_{i,k}K_{\bot}s_{ik}s_{ik+1},
\end{equation}
where $s_{ik}$ denotes the classical Ising spins with values $\pm1$,  and $k$ is the index in the imaginary time direction of the (2+1)-D classical system. It consists of antiferromagnetic coupling $K_{ij}=J/(nT)$ within each layer and ferromagnetic coupling $K_{\bot}=(1/2)\ln(nT/\Gamma)$ between layers. The map of quantum to classical system becomes exact when the Trotter number $n$ goes to infinity. Thus, it allows us to study the transverse field TIAF numerically, e.g.,  by means of continuous time Monte Carlo algorithm~\cite{rieger}, which avoids the exponential increasing of the system's height as $\exp(2K_{\bot})$. Instead of discretizing the imaginary time direction and storing values of $\pm1$ at each Trotter step, we take the continuous time limit for applying the scheme of the Swendsen-Wang cluster update method~\cite{wang}. We cut a continuous segment $\tilde{S}_i\{[t_0,t_0+t]\}$ of length $t<\beta$ by a Poisson process with decay rate of $1/\Gamma$. Next we connect space-neighboring segments $\tilde{S}_i\{[t_1,t_2]\}$ and $\tilde{S}_j\{[t_3,t_4]\}$ with an overlap length of $t=\text{len}([t_1,t_2]\cap[t_3,t_4])$ by the probability of $1-\exp(-2Jt)$. This way,  we build clusters of connected segments, randomly assigning them a value of $\pm1$, removing unessential cuts in each segments for obtaining a new configuration of the $(2+1)$-D system. We update the system by Metropolis acceptance criterion $p=\min{1,\exp(-\Delta E)}$, where $\Delta E$ denotes the energy difference between the original and the new configuration. We retrieve the local magnetization $m_i$ by dividing the weighted length of the spin $\sigma_i$ by $\beta$, from which we can detect the phase of the transverse field TIAF with arbitrary choice of $T$, $J$, and $\Gamma$. The details of the algorithm are explained in Appendix.~\ref{sec:MC}.

\subsection{Problem formulation}
The analytical description and the computational approach provide us the tools to study the exotic phases of the magnets with geometrical frustration and phase boundaries. We assume that Bob does not have enough prior knowledge of the system, i.e., Bob only knows that there exist three distinguishable phases in the transverse field TIAF model, which are a paramagnetic (disordered) phase at high temperatures or strong transverse fields, a clock (ordered) phase at low temperature or weak transverse field, and a floating KT (critical) phase between them. Bob aims at predicting the phase boundaries by ML models, with samples be $\mathbf{x_i}=(\Gamma,T,J)$. Bob can detect the phase of the model under the parameter $\mathbf{x_i}$ by either computational physics approach or condensed matter physics experiments, labeling the sample by the outcome $y_i$. Considering that labeling samples by either method might be time-demanding and expensive, Bob employs AL for an optimal experiment design, labeling the sample with maximum uncertainty evaluated by the model. SSL comes behind the AL after labeling as many samples as possible, corresponding to the experiment's budget for artificial labeling.

\subsection{Numerical simulation}
Instead of sampling on a lattice with finite qutrits, Bob is facing a continuous parameter space for sampling, which can be simplified after discretization of the space. Since Bob already knows that there is a KT phase floating between the others, Bob selects a nonlinear support vector machine (SVM) with a Gaussian kernel. Although SVM is not a probabilistic model, one can still equivalently evaluate the uncertainty by its decision function. Accordingly, Bob transforms the triple classification problem to binary classification by combining two one-vs-rest (OvR). Different phases are characterized by positive or negative values of the decision function. It is trivial that the estimations of the phase boundaries are the counter lines of the value to be zero. In order to evaluate the performance of AL and its combination with SSL, one has to define the correct rate by comparing the values of the decision function to the true value. We highlight that even though the analytical analysis leads us to an elegant expression of phase boundaries, as shown in eq.~\eqref{eq:boundary}, the parameters $b$, $\nu$, and $\Gamma_c$ are still suggested by previous numerical studies instead of calculating ab initio. In other words, these parameters have their uncertainties, which are bounded by shots of DMRG or MC numerical experiments. Thus, Alice assumes that the combination of analytical expressions and experimental values could characterize the phase boundaries, i.e., Alice employs Eq.~\eqref{eq:boundary} as the classifier for obtaining true labels of samples for model evaluation, which might not exactly describe the phase boundaries in nature. Meanwhile, various computational physics approaches with hyperparameters result in different phase-detection outcomes, especially when the sample is near the true phase boundaries. Thus, Bob queries the phase of a sample from Alice instead of doing numerical simulations for a fair evaluation of the AL algorithm. To mimic the statistical error in numerical or laboratory experiments, Alice models the error by a probability of wrong labeling, flipping the oracle's OvR class by
\begin{equation}
\label{eq:error}
P_{\text{flip}}=\frac{1}{2}\exp(-kd),
\end{equation}
where $k$ is a tunable coefficient and $d$ is the distance from the sample site in parameter space to the curve.

\begin{figure}
\includegraphics[width=7.8cm]{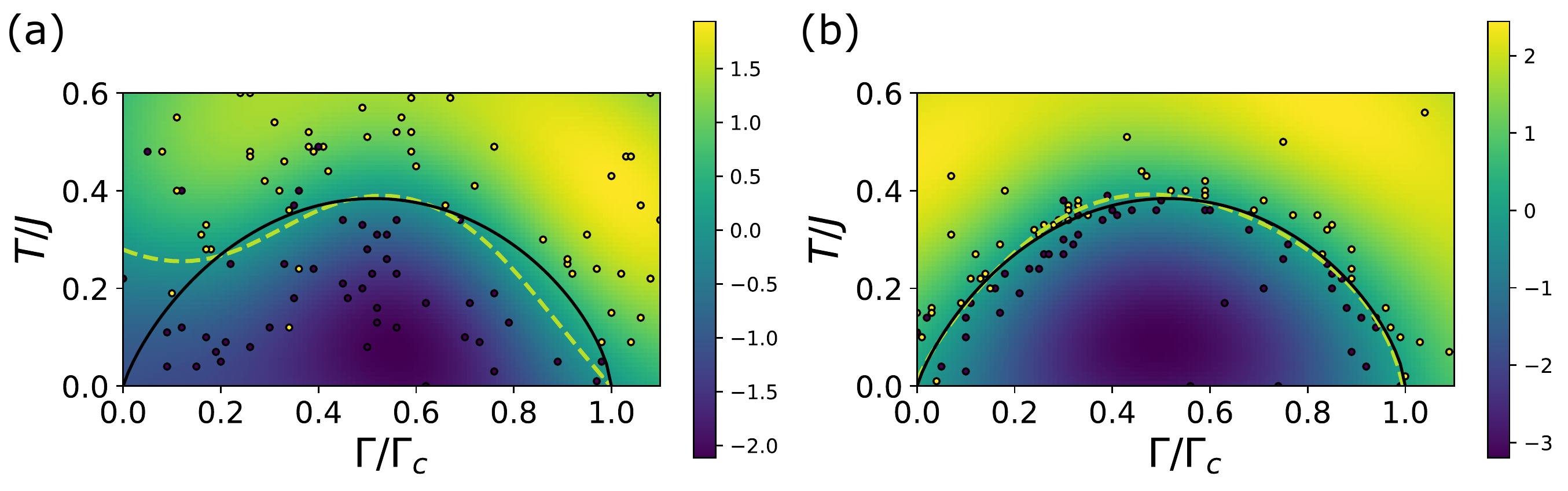}
\caption{\label{fig:svmrslc}(a) A binary classification model with sampling strategy be random sampling, discriminating the paramagnetic phase and rest. We show the heatmap of the decision function, and plot the $100$ queried samples by their labels. Labeling error is modeled by Eq.~\eqref{eq:error} with $k=50$. By labeling less than 2\% of the samples, the model achieves a rate estimation of 94\%. (b) The strategy is margin sampling which improves the performance to 99\% while the other parameters remain the same.}
\end{figure}

\begin{figure}
\includegraphics[width=7.8cm]{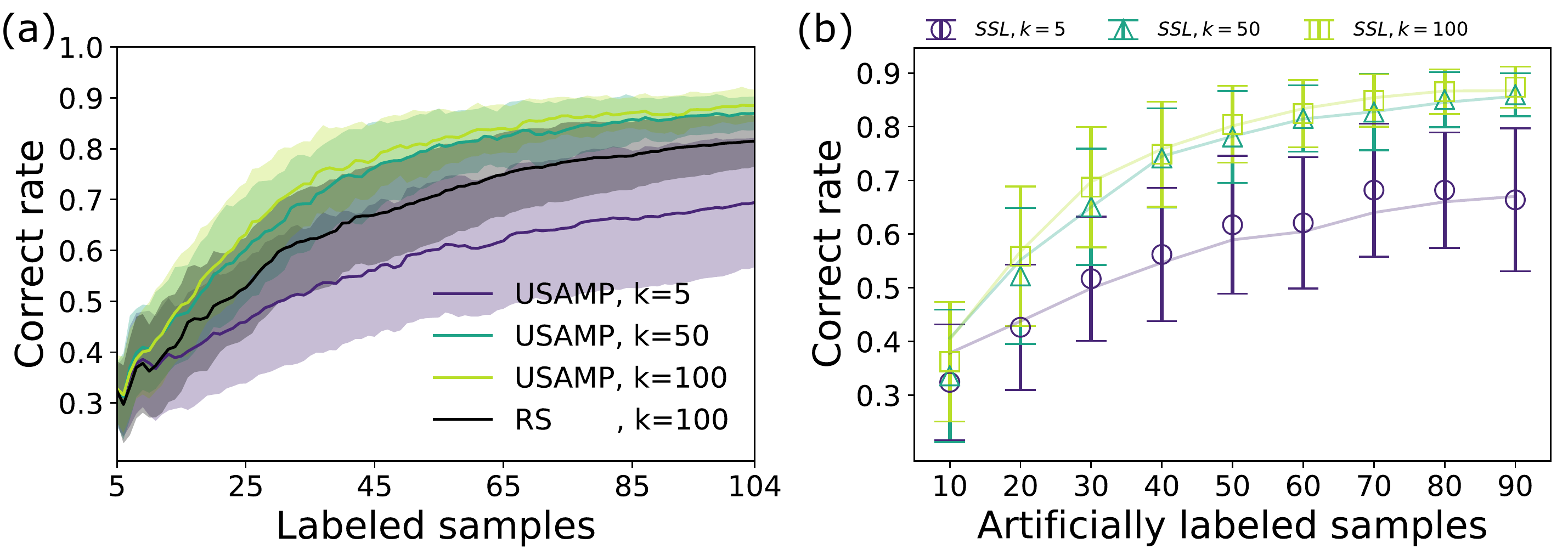}
\caption{\label{fig:ssl} (a) Mean correct rates of estimation on phase boundaries with random sampling (baseline), USAMP (k=5, 50, 100) as different sampling strategies and labeling error rates. Confidence intervals are filled by transparent colors, denoting a standard deviation based on 100 numerical experiments. (b) Mean correct rates of AL-trained models followed by SSL, after the numbers of samples labeled by annotator meet the upper limits. The transparent lines denote the mean correct rate of AL-trained models without SSL. Parameters remain the same as in the previous subfigure.}
\end{figure}

In Fig.~\ref{fig:svmrslc}, we present a demonstration on predicting the boundary of the paramagnetic phase and the rest by the use of RS as baseline and USAMP. We notice that USAMP queries samples close to the analytical phase boundary, achieving a better estimation than RS does. Then we go for a quantitative study by combining two nonlinear SVM with Gaussian kernel for discriminating among three quantum phases. We set the space to be sampled as $\Gamma/\Gamma_c\in[0,1.1]$, $T/J\in[0,0.6]$ for paramagnetic phase vs rest and $T/J\in[0,0.3]$ for ordered phase vs rest, respectively, where the lattice length is $0.01$. We benchmark the performance of USAMP under different labeling error rates by RS [c.f. Fig.~\ref{fig:ssl}(a)]; USAMP with labeling error modeled by $k=100$ achieves almost 90\% rate estimation for the triple classification problem, outperforming RS significantly. We also perceive the fact that the RS with $k=100$ realizes a more precise estimation than the USAMP with $k=5$ does. We deduce that it is reasonable because $k=5$ has a far higher probability of incorrect labeling of samples. To be more specific, there is a trade-off when labeling the samples that are close to the phase boundaries, because they reduce uncertainty but the model is more likely to be misled by introducing wrong labels. By contrast, RS uniformly selects samples, which most of them are far from the boundaries. Although labeling them does not reduce the uncertainty as much as USAMP does, the possibility of incorrect labeling is notably suppressed, especially when the parameter $k$ is small enough in our error model. We further consider the situation that the experimental budget is very limited, i.e., one can only label minimal numbers of samples, requiring the cooperation between AL and SSL. In Fig.~\ref{fig:ssl}(b), we use SSL to investigate if one can refine the model trained by AL once the number of artificially labeled samples meets the upper limit. We verify that SSL hardly improves the performance of the model trained by AL. Therefore, we conclude that AL provides an optimal design of experiments, exploiting the information in each queried sample.

\section{Conclusions}
We have proven the advantage of AL algorithms on studying multinomial classification problems in physics by two representative cases, namely, quantum information retrieval in qutrits and phase boundary prediction in many-body physics. AL algorithms remarkably reduce the cost of labeling, which can be defined theoretically or practically, by fidelity loss in quantum information retrieval or operation cost in physics experiments, respectively. The AL-assisted information retrieval in qutrits can be experimentally verified by entangling photons, with biphoton locally projected by Alice and remotely prepared at Bob, followed by two ancilla photons for extracting information. AL algorithms enable optimal model-training, thereby equivalently quantifying the relation between quantum fidelity loss and the corresponding classical information gain by performance enhancement of the model without calculating the information entropy reduction~\cite{gain}. One should also focus attention on the fact that the performance of AL is also related to the ML model itself. Although it significantly reduces the labels required for model training for most tasks, it fails when the ML model does not work for given data distribution or even misleads the prediction, with the counterexample shown in the Appendix. 

A possible extension of this work could involve the combination of the proposed framework with deep learning. An artificial neural network (ANN), as a universal functional approximator, may substitute the statistical ML models as the discriminator. Pre-trained ANNs can also work as generative models, fundamentally related to the origin of trade-off and dynamical prediction~\cite{gm}, or annotating samples for effectively training a less complex model with more interpretability~\cite{interpretable}. The difficulty of this extension is that repetitively training ANNs with updating datasets costs too many computational resources, violating the aim of efficient training by itself. Such extension requires training theory for the dynamical dataset from the community of computer science. 

Another extension is multi-label classification, which should not be confused with the multinomial classification presented in this work. Multi-label classifications involve that multiple labels can be predicted for each sample, which naturally allows the encoding in quantum systems by superpositions. 

One more possibility to enhance the research presented in this work is to include information about data density when selecting the labels to be sampled in order to avoid choosing samples that are close to the decision border but are not representative of the data set.

Finally, we would like to emphasize that the use of AL in physics is apparently not restricted to retrieving quantum information and studying many-body physics. AL algorithms assist physicists in designing optimal experiment strategies and analyzing the data output, leading to promising applications in particle physics and cosmology as well, where the cost of the experiments is extremely sensitive.

\begin{acknowledgements}
This work is partially supported from NSFC (12075145), STCSM (2019SHZDZX01-ZX04 and 20DZ2290900), SMAMR (2021-40), Program for Eastern Scholar, QMiCS (820505) and OpenSuperQ (820363) of the EU Flagship on Quantum Technologies,  EU FET Open Grant Quromorphic, Spanish Government PGC2018-095113-B-I00 (MCIU/AEI/FEDER, UE), and Basque Government IT986-16. X. C. acknowledges Ram\'on y Cajal program (RYC-2017-22482).

\end{acknowledgements}

\begin{appendix}

\section{Machine learning models}
We briefly introduce the ML model for optimal training with a few labeled samples: logistic regression, nonlinear SVM with Gaussian kernel, and naive Bayes. Semi-supervised learning and its implementation are also provided for reproduction. All the algorithms are available in the latest version (v.0.24.2) of \textsc{Python} library \textsc{scikit-learn}~\cite{scikit}.

\textit{Logistic regression.---} Logistic regression is a linear model for classification, which is also known as maximum-entropy classification. The binary prediction of a sample can be modeled by the logistic function
\begin{eqnarray}
P_{\theta}(y_i=1|\mathbf{x_i})&=&h_\theta(\mathbf{x_i})=\frac{1}{1+\exp(-\theta^{\text{T}}\mathbf{x_i})},\\
P_{\theta}(y_i=0|\mathbf{x_i})&=&1-P_{\theta}(y_i=1|\mathbf{x_i}),
\end{eqnarray}
where the generalized linear separation model reads
\begin{equation}
\log\left(\frac{y}{1-y}\right)=\theta^{\text{T}}\mathbf{x}=\theta_0+\theta_1x_1+\cdots+\theta_nx_n.
\end{equation}
For the model training, we write down the cost function to be minimized as a likehood function, 
\begin{equation}
\label{eq:logistic}
J(\theta)=-N^{-1}\sum_{i=1}^N\log P_\theta(y_i|\mathbf{x_i}),
\end{equation}
assuming that all observations in the sample follow the Bernoulli distribution. The gradient of the cost function gives
\begin{equation}
\partial_{\theta_j}J(\theta)=N^{-1}\sum_{i=1}^N\left[h_\theta(\mathbf{x_i})-y_i\right]x_j,
\end{equation}
which updates the parameter by $\theta_j=\theta_j-\alpha\partial_{\theta_j}J(\theta)$. It is equivalent to summing the conditional entropy and Kullback-Liebler divergence in the limit of large $N$. For finding the best $h\theta(\mathbf{x})$, $\theta^{\text{T}}\mathbf{x}$ will be as small (large) as possible for producing class label $y=0$ (or $y=0$), thereby let $\theta_j\rightarrow-\infty$ (or $+\infty$). To avoid such overfitting and numerical instability for a better generalization, one considers controlling the growth of $\theta_j$ by regularization. A typical $L2$ regularization penalizes the cost function by rewriting the cost function in the following way
\begin{equation}
\tilde{J}(\theta)=J(\theta)+\frac{\lambda}{2N}\theta^{\text{T}}\theta,
\end{equation}
where $\lambda$ is the hyperparameter that leads to underfitting by making $\theta_j$ shrink to $0$ with a large value and has less regularization effect with a smaller value. Logistic regression is a vital ML model that inspires famous extensions for other tasks, e.g., conditional random field in natural language processing, aiming to learn the sequential features of samples.

In our implementation with \textsc{scikit-learn}, we use $L2$ regularization with hyperparameter $C=1/\lambda=1$ for a multinomial classification of quantum information retrieval in qutrits. The model is trained by Limited-memory Broyden-Fletcher-Goldfarb-Shanno (LBFGS) algorithm for gradient descent, where learning rate, maximum iteration, and tolerance are set to default.

\textit{Nonlinear Gaussian SVM.---}
Support vector machine aims at finding a hyperplane that separates support vectors $\mathbf{x_i}$ (samples) of different classes $y=\pm1$ with the correct prediction for most support vectors. For each sample we define a variable $\xi_i=\max\left[0,1-y_i(w^{\text{T}}\mathbf{x_i}-b)\right]$. Equivalently, we highlight that $\xi_i$ is the minimal nonnegative value satisfying $y_i(w^{\text{T}}\mathbf{x_i}-b)\geq1-\xi_i$. To this goal, we solve the primal problem
\begin{equation}
\min_{w,b,\xi_i} \frac{1}{2}w^{\text{T}}w+C\sum_{i=1}^n\xi_i,
\end{equation}
where $C$ is the $L2$ regularization parameter for controlling the distances from the samples to the hyperplane. One derives the Lagrangian dual problem to the primal as
\begin{equation}
\max_{c_i}=-\frac{1}{2}\sum_{i=1}^N\sum_{j=1}^Ny_ic_i(\mathbf{x_i}^{\text{T}}\mathbf{x_j})y_jc_j,
\end{equation}
where $\sum_{i=1}^Nc_iy_i=0$ and $0\leq\ c_i\leq C$. For a nonlinear classification, one may map the support vectors into a higher dimensional space as $\mathbf{x_i}\rightarrow\phi(\mathbf{x_i})$. The inner product $\mathbf{x_i}^{\text{T}}\mathbf{x_j}$ is substituted by the kernel $K(\mathbf{x_i},\mathbf{x_i},\mathbf{x_j})=\phi(\mathbf{x_i})\phi(\mathbf{x_j})$.

For the phase boundary prediction problem, we use the radial basis function (Gaussian) kernel
\begin{equation}
K(\mathbf{x_i},\mathbf{x_j})=\exp\left(-\gamma||\mathbf{x_i}-\mathbf{x_j}||^2\right),
\end{equation}
where $||\mathbf{x_i}-\mathbf{x_j}||$ is the Euclidean distance between $\mathbf{x_i}$ and $\mathbf{x_j}$. Once we solve the optimization problem, we obtain the decision function for a given sample
\begin{equation}
\sum_i y_ic_iK(\mathbf{x_i},x)+b,
\end{equation}
where a positive (negative) value denotes the label prediction of $y=\pm1$. In our implementation, we set $C=1$ and $\gamma=1/[2\text{var}(X)]$.

\textit{Naive Bayes.---}
Naive Bayes is a simple ML model for predicting the class of a sample $\mathbf{x_i}$ by conditional probabilities
\begin{equation}
P(y|\mathbf{x_i}) = \frac{P(y)P(\mathbf{x_i}|y)}{P(\mathbf{x_i})}.
\end{equation}
The naive conditional independence assumption
\begin{equation}
P(x_j|y,x_1,\cdots,x_{j-1},x_{j+1},\cdots,x_n)=P(x_j|y)
\end{equation} 
simplifies the prediction as
\begin{equation}
P(y|\mathbf{x_i})=\frac{P(y)\Pi_{i=1}^nP(x_j|y)}{P(\mathbf{x_i})}.
\end{equation}
Thus, one realizes the classification by following the rule
\begin{equation}
\hat{y}=\underset{y}{\text{argmax}}P(y)\Pi_{j=1}^{n}P(x_j|y),
\end{equation}
where naive Bayes classifiers differ by the assumption of $P(x_j|y)$. In our extra AL numerical experiments, we assume that the distribution is Gaussian
\begin{equation}
P(x_j|y)=\frac{1}{\sqrt{2\pi\sigma_y^2}}\exp\left[-\frac{(x_j-\mu_y)^2}{2\sigma_y^2}\right].
\end{equation}

\textit{Semi-supervised learning.---} In our implementation, we use a self-training classifier to learn from unlabeled data. After training a nonlinear Gaussian SVM by AL, the SVM iteratively predicts pseudo-labels for all data in the unlabeled pool. A sample is added to the training set with its pseudo-label once the confidence meets a threshold. The SVM trains itself until the maximum iteration number is met, or no new samples were added in the last iteration. For the phase boundary prediction problem, we set the threshold to $0.95$, and the maximum iteration number is $5$.

\section{\label{sec:AL}Extra numerical experiments in qutrits}
Here we present miscellaneous information about the extra numerical experiments on quantum information retrieval in qutrits. Besides the dataset described in the main text (\textit{Case I}), we test AL algorithms on another configuration without rotational symmetry (\textit{Case II}). We introduce the function to generate the dataset as follows.

\textit{Case I.---} We define the rotation operation $\mathbf{x'}=R(\theta)\mathbf{x}$ on the parameter space $\mathbf{x}=(x_1,x_2)^{\text{T}}\in[-1,1]\times[-1,1]$
\begin{equation}
\begin{pmatrix}
x_1' \\x_2'
\end{pmatrix}=\begin{pmatrix}
\cos\theta & -\sin\theta \\ \sin\theta & \cos\theta
\end{pmatrix}
\begin{pmatrix}
x_1 \\ x_2
\end{pmatrix}.
\end{equation}
We have three independent parameter space $\mathbf{x^{(1)}}=R(0.32)\mathbf{x}$, $\mathbf{x^{(2)}}=R(2\pi/3+0.32)\mathbf{x}$, and $\mathbf{x^{(3)}}=R(4\pi/3+0.32)\mathbf{x}$, giving angular parameters $\phi^{(i)}=\arctan(x^{(i)}_2/x^{(i)}_1)$. Thus, we generate the dataset by
\begin{eqnarray}
\tilde{c}_i(x_1,x_2)&=&\frac{1}{2}\left(1+\sin\phi^{(i)}\right),~x^{(i)}_1 \geq 0,\nonumber\\
~&=&\frac{1}{2}\left(1-\sin\phi^{(i)}\right),~x^{(i)}_1<0,
\end{eqnarray}
where the amplitudes $c_i$ are derived from the normalization of $\tilde{c}_i$.

\textit{Case II.---} Here the parameter space is defined on $\mathbf{x}=(x_1,x_2)^{\text{T}}\in[0,\pi/4]\times[0,\pi/4]$, where amplitudes of each basis reads
\begin{eqnarray}
c_1(x_1,x_2)&=&\sin^2(x_1+x_2),\nonumber\\
c_3(x_1,x_2)&=&\cos^2(x_1+x_2),\nonumber\\
c_2(x_1,x_2)&=&\sqrt{\left|1-c_1^2-c_3^2\right|}.
\end{eqnarray}

\begin{figure}
\includegraphics[width=7.8cm]{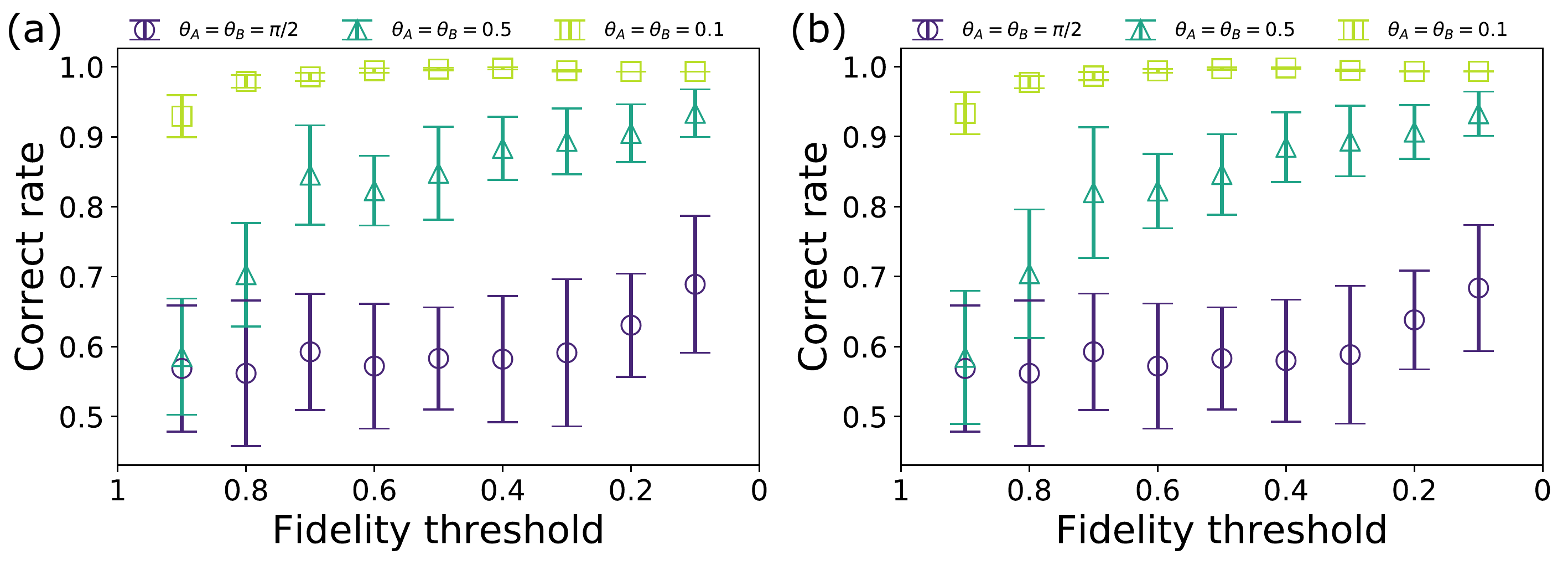}
\caption{\label{fig:suplces} Mean correct rates of logistic regression triple classification model for \textit{Case I} with sampling strategies as least confidence (a) and entropy sampling (b), respectively. Each qutrit is labeled by measurements of different strength $\theta_A$ and $\theta_B$. Parameters remain the same as in the previous subfigure.}
\end{figure}

\textit{Benchmarking sampling strategies.---} In the main text, we demonstrate the performance of AL algorithms with different sampling strategies for training the logistic regression model. For \textit{Case I},  margin sampling gives a smoother curve of correct rate [c.f. Fig.~\ref{fig:fidelity}(a)]. Here we present the correct rate of least confidence and entropy sampling for the same test in Fig.~\ref{fig:suplces}. As we expected, the performances are hardly distinguishable.

\textit{Naive Bayes for Case I.---} In the main text, we mentioned that one could not take advantage of AL for granted. Sometimes the baseline RS outperforms all USAMP methods since AL cooperates with the model, whose mechanism affects the sampling behavior.

\begin{figure}
\includegraphics[width=7.8cm]{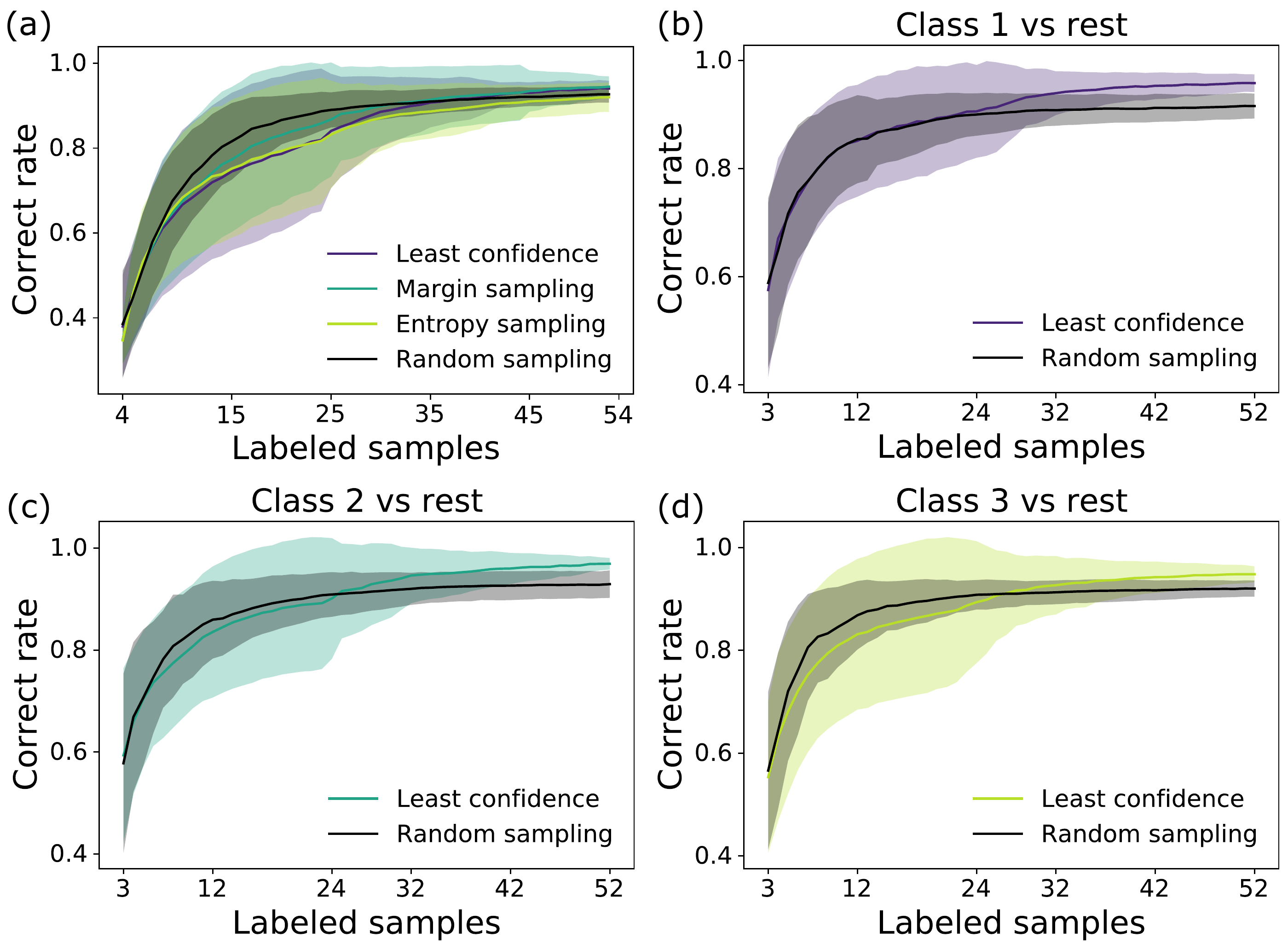}
\caption{\label{fig:supnb1} (a) Mean correct rates of naive Bayes triple classification model for \textit{Case I} with random sampling (baseline), least confidence, margin sampling, and entropy sampling as different sampling strategies. Confidence intervals are filled by transparent colors, denoting a standard deviation based on $200$ numerical experiments. (b-d) Mean correct rates of class 1, 2, 3 versus rest, respectively, where USAMP strategies are equivalent for binary classification problems. The other parameters remain the same.}
\end{figure}

Here we test AL algorithms to efficiently train Naive Bayes classifiers, aiming to solve the quantum information retrieval in qutrits. In \textit{Case I}, we proved that the logistic regression model gives a satisfying prediction with USAMP, and significantly outperforms RS. However, we meet a setback in training a Naive Bayes classifier for the same task (c.f. Fig.~\ref{fig:supnb1}). We test OvR for three classes; although the performance of USAMP surpasses RS with more labeled samples, both show similar behaviors. We reckon that the similar performances are reasonable since the quantum information dataset is generated with rotational symmetric so that each class is balanced. As a result, RS outperforms all USAMP strategies with a small number of labeled samples.

\begin{figure}
\includegraphics[width=7.8cm]{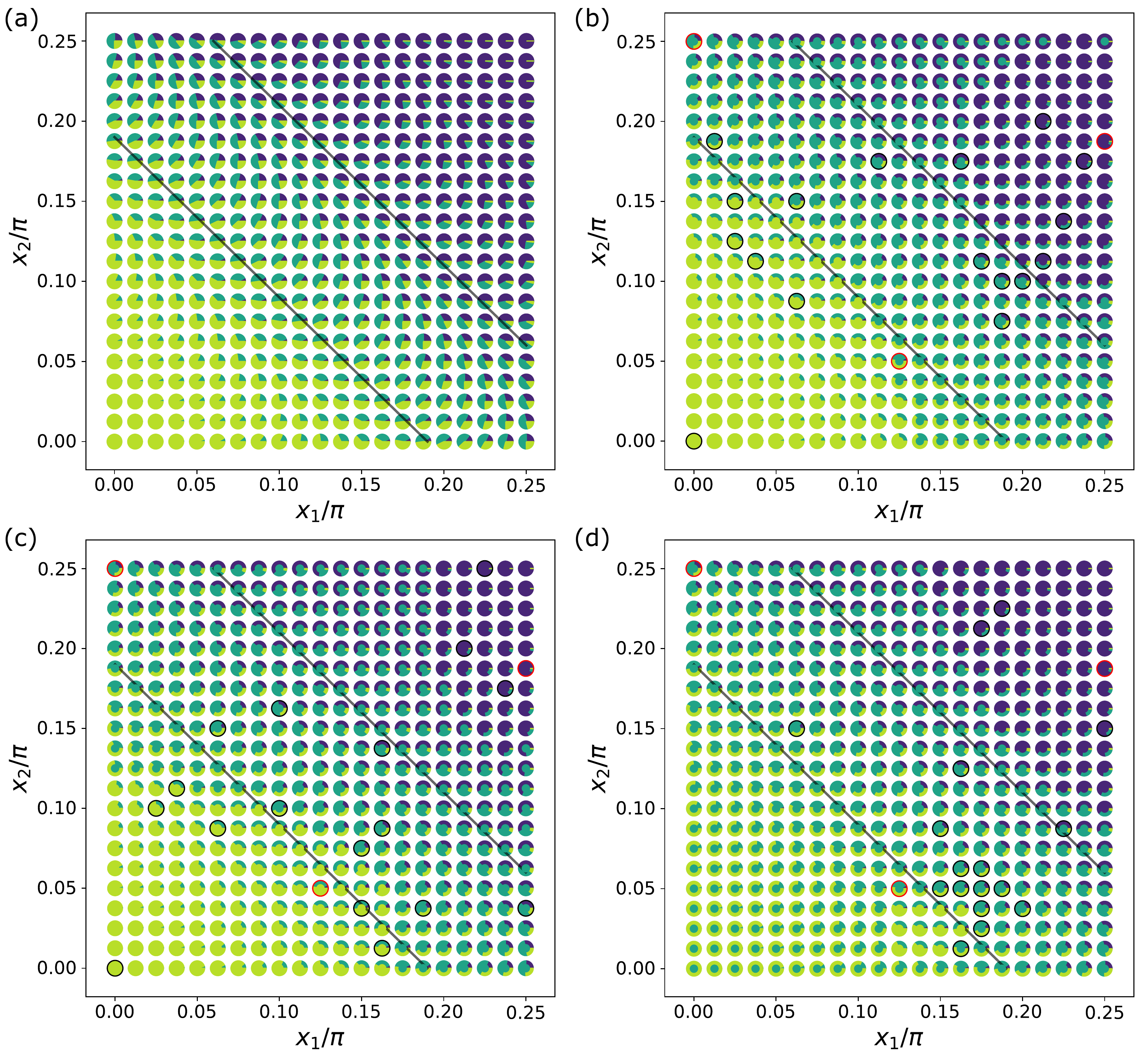}
\caption{\label{fig:supalicebob} (a) The lattice prepared by Alice, encoding the quantum information of \textit{Case II}, which consists of $21\times21=441$ qutrits for triple classification. The qutrit states are demonstrated by pie charts with proportions be $|c_1|^2$, $|c_2|^2$, and $|c_3|^2$. (b), (c), and (d) USAMP protocols with the query strategies be least confidence, margin sampling, and entropy sampling, respectively. Bob initializes a naive Bayes model by three oracles provided by Alice (circled by red). Qutrits queried by the model are circled by black and covered by smaller circles in class colors. }
\end{figure}

\begin{figure}
\includegraphics[width=7.8cm]{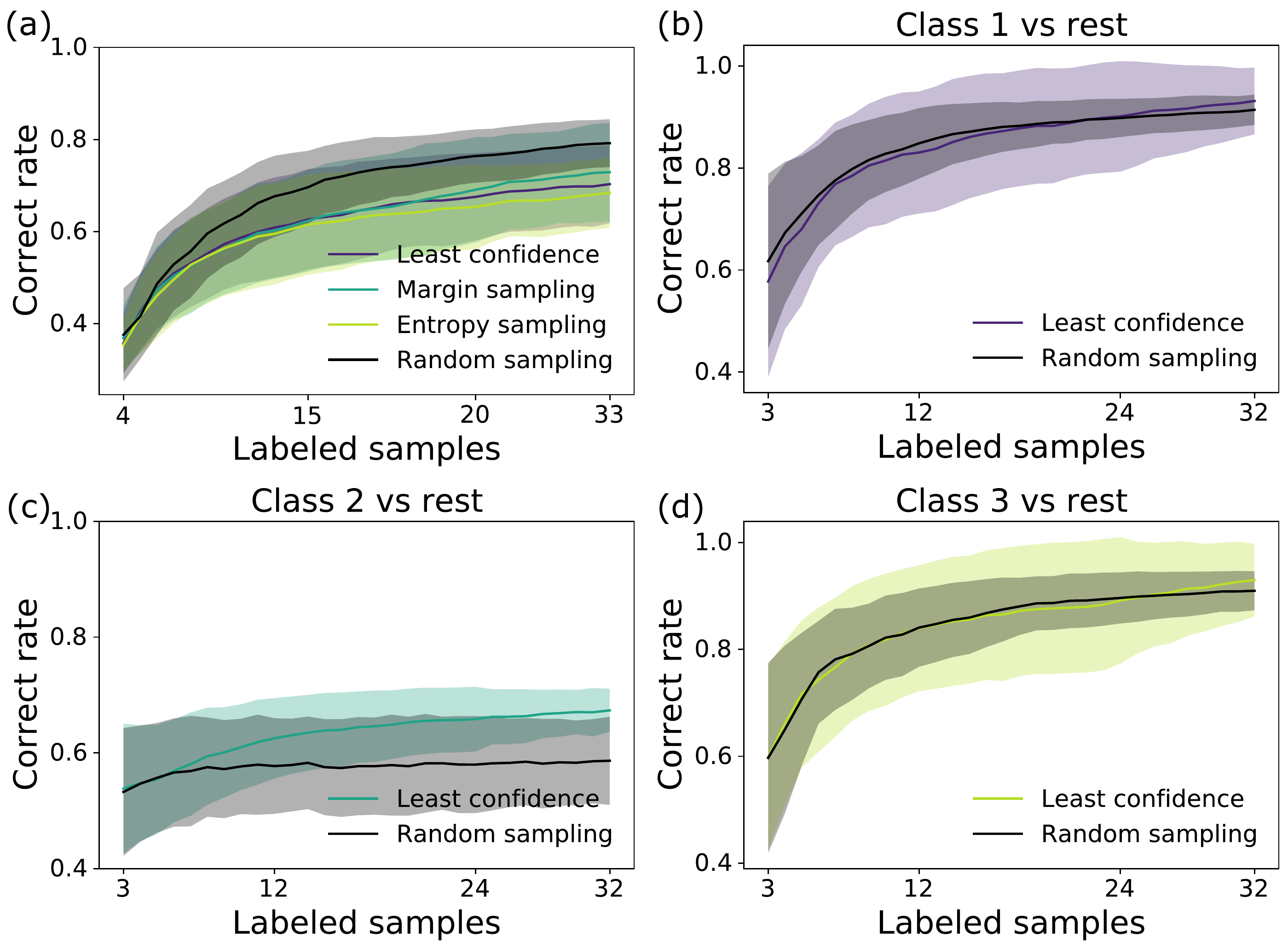}
\caption{\label{fig:supnb2} (a) Mean correct rates of naive Bayes triple classification model for \textit{Case II} with random sampling (baseline), least confidence, margin sampling, and entropy sampling as different sampling strategies. Confidence intervals are filled by transparent colors, denoting a standard deviation based on $200$ numerical experiments. (b-d) Mean correct rates of class 1, 2, 3 versus rest, respectively, where USAMP strategies are equivalent for binary classification problems.  The other parameters remain the same.}
\end{figure}

\textit{Naive Bayes for Case II.---} We wonder if the phenomenon in \textit{Case I} exists in other scenarios, which inspired us to test the naive Bayes classifier on another dataset as \textit{Case II}. We present an illustrative demonstration on the query behaviors and a quantitative study in Fig.~\ref{fig:supalicebob} and~~\ref{fig:supnb2}, respectively. We find out that RS outperforms USAMP strategies even more significantly in \textit{Case II} than in \textit{Case I}.

Now we analyze the mechanism of the fail of AL algorithms in such model and dataset. AL algorithms collaborate with ML models, querying the most informative sample according to the model's estimation. It might be misled if the probabilistic model is based on the prior, which is sensitive to the labels of samples in the training set. Indeed, the predictions of unknown samples via naive Bayes classifier are in proportion to $P(y)$ as the prior. For example, the model will ignore the probability of a third class, if the samples of that third class do not share enough population in the preliminary training set, with only a few samples. Thus, the model performs an almost binary classification with AL, which can only be corrected once almost all samples are queried. By contrast, RS queries samples uniformly distributed in the parameter space. Although the samples queried are supposed to be less informative, it ensures a better estimation of $P(y)$ by querying samples that uniformly distribute in the parameter space. 


\section{\label{sec:MC}Continuous time Monte Carlo algorithm}
For fairly evaluating the ML models with true values, we do not employ the continuous time Monte Carlo algorithm for labeling the samples by detecting its phase, but simulating a numerical error model corresponding to the analytical analysis. We reckon it is necessary to introduce the algorithm for a better understanding, which might help the audience apply AL algorithms to other tasks in many-body physics, where one aims to predict unknown features of many-body systems, i.e., model evaluation is no longer required.

For a reliable Monte Carlo approach, one has to verify if the sign problem exists in the system to be studied. There are only positive weights in the transverse TIAF model, that is to say, there is no sign problem for this geometrically frustrated system. According to Suzuki-Trotter's theorem~\cite{suzuki}, the 2D quantum Ising model is equivalent to (2+1)-D classical Ising model, whose discretized version is difficult to be simulated because of the scaling limit. Thus, one can use a continuous (imaginary) time algorithm~\cite{rieger} that no longer treats classical spin variables as lattices, but continuous segments of a certain length: $s_{ik}=s_{ik+1}=\cdots=s_{ik+N}\rightarrow\tilde{S}_i\{[t_0,t_0+t]\}$, where $t=N/(nT)$. The edges of the segments are named cuts, which at those switching times, the spin
is flipped to another value.

\begin{figure}
\includegraphics[width=7.8cm]{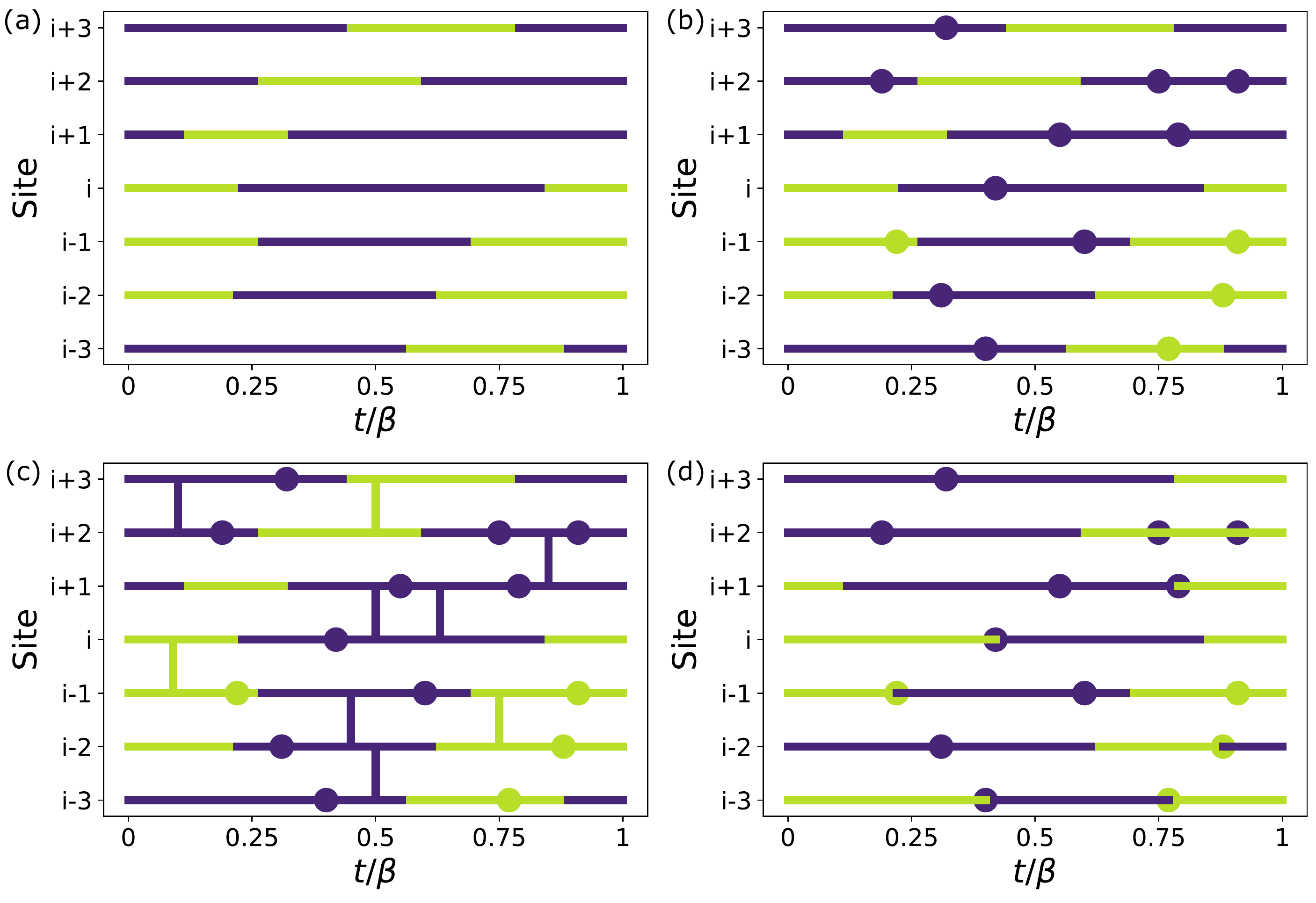}
\caption{\label{fig:sw} (a) An initial configuration of the system, consists of spin segments in the imaginary time direction. (b) Inserting new cuts according to the Poisson process~\eqref{eq:poisson}. (c) Connecting segments for clustering~\eqref{eq:clustering}. (d) New configuration after randomly assigning values to each cluster. The redundant cuts within segments should be removed before a new iteration.}
\end{figure}

The configuration is updated by the Swendsen-Wang method, consisting of inserting new cuts and clustering segments as two main steps. For neighboring classical spins on the imaginary time direction $s_{ik}$ and $s_{ik+1}$, the probability of sharing a same value is $p_i=1-\exp(-2K_\bot)=1-(\Gamma/nT)+O(1/n^2T^2)$. Accordingly, the probability of connecting classical spins on the imaginary time direction of a length $t$ in the continuous limit reads
\begin{equation}
\label{eq:poisson}
(p_i)^{ntT}=(1-\Gamma/nT)^{ntT}\rightarrow\exp(-\Gamma t).
\end{equation}
It denotes a Poisson process with decay time $1/\Gamma$. Similarly, we connect space-neighboring segments $\tilde{S}_i\{[t_1,t_2]\}$ and $\tilde{S}_j\{[t_3,t_4]\}$ with an overlap length of $t=\text{len}([t_1,t_2]\cap[t_3,t_4])$ by the probability
\begin{equation}
\label{eq:clustering}
1-(1-p_{ij})^{ntT}=1-(1-2K_{ij}/nT)^{ntT}\rightarrow1-\exp(-2K_{ij}t),
\end{equation}
forming new clusters for a value assignation of $\pm1$ with equal probability. A schematic of the cluster-updating method is shown in Fig.~\ref{fig:sw}.

It is straightforward to obtain the measurement of observables, e.g., the local magnetization $m_i$ can be calculated by averaging the length of segments with weights be $\pm1$, and then divided by $\beta=1/T$, leading to local susceptibility
\begin{equation}
\chi_i=\int_0^\beta\langle\sigma_i^z(\tau)\sigma_i^z(0)\rangle d\tau=\beta\langle m_i^2\rangle,
\end{equation}
by averaging over configurations generated over Monte Carlo steps. For labeling samples, i.e., detecting phases, we briefly introduce the quantities for completing such task. Landau-Ginzberg-Wilson theory suggests that the complex XY order parameter~\cite{lgw} is
\begin{equation}
\psi(\mathbf{r})=\psi_0\exp(i\phi)=\left[m_1+m_2\exp\left(\frac{i4\pi}{3}\right)+m_3\exp\left(-\frac{i4\pi}{3}\right)\right]/\sqrt{3},
\end{equation}
which equals to zero for paramagnetic phase and critical phase, where $m_i$ are the local magnetizations of spins in the triangular sublattice. One can also calculate the susceptibility once the magnetization vanishes in the limit of infinite large TIAF. Clock phase can be detected by checking the sixfold symmetry breaking term
\begin{equation}
c_6=\frac{\langle\psi_0^6\cos6\theta\rangle}{\langle\psi_0^6\rangle},
\end{equation}
which equals to $\pm1$ for $(+--)$ or $(+0-)$ phase, as well as $0$ for paramagnetic phase and critical phase. Another quantity that distinguishes paramagnetic phase and critical phase is the Binder cumulant
\begin{equation}
U=1-\frac{\langle\psi_0^4\rangle}{3\langle\psi_0^2\rangle^2},
\end{equation}
which converges to 0 at paramagnetic phase, $\tilde{U}$ that depends on the location within critical phase, and $2/3$ at clock phase in the infinite large system size limit. Locating the critical phase can also be approached by scaling analysis~\cite{sixfold} and KT theory~\cite{kt}, which is introduced in the numerical study of the model with full details~\cite{isakov}.

\end{appendix}

\end{document}